\documentclass[usenatbib]{mn2e}                                                 
\usepackage{graphicx}                                                           
\usepackage{aas_macros}                                                         
\usepackage{amsfonts}                                                           
\usepackage{bm}  %allows bold fonts for Greek letters                           
\usepackage{times}                                                              
\begin{document}                                                                
                          
% Extra definitions                                                                                
\def\beqn{\begin{equation}}
\def\eeqn{\begin{equation}}
\def\beqnarray{\begin{eqnarray}}                                                
\def\eeqnarray{\end{eqnarray}}                                                  
\def\norml{{\bf{\hat{l}}}}                                                      
\def\norms{{\bf{\hat{s}}}}                                                      
\def\vecv{{\bf{v}}}                                                             
\def\deg{\hbox{$\null^\circ$}}                                                  
                                  
\def\Msun{\,\textrm {M}_\odot}                                              
\def\kpc{\,{\rm kpc}}                                                           
\def\Gyr{\,{\rm Gyr}}                                                           
\def\Myr{\,{\rm Myr}}                                                           
\def\kms{\,{\rm km}\,{\rm s}^{-1}}
\def\cm2{\,\textrm {cm}^{-2}}                                      
                   
\def\I{\textrm{\scriptsize{I}}}
\def\sI{\textrm{\small{I}}}
\def\CL{\textrm{\small{CL}}}
\def\HVC{\textrm{\small{HVC}}}
\def\CHVC{\textrm{\small{CHVC}}}
\def\IC{\textrm{\small{IC}}}
\def\ud{\mathrm{d}}                                                             

\title[GCN: a gaseous Galactic halo stream?]{GCN: a gaseous Galactic halo stream?}

\author[S.~Jin]
{Shoko Jin\thanks{e-mail: shoko@ari.uni-heidelberg.de; Alexander von Humboldt research fellow}$^{1,2}$\\
$^1$Astronomisches Rechen-Institut, Zentrum f\"ur Astronomie der Universit\"at Heidelberg, M\"onchhofstr. 12--14, D-69120, Heidelberg, Germany\\
$^2$Institute of Astronomy, University of Cambridge, Madingley Road, Cambridge, CB3 0HA, U.K.}
\date{}

\maketitle

\begin{abstract}

We show that a string of H$\I$ clouds that form part of the high-velocity cloud complex known as GCN is a probable gaseous stream extending over more than $50\deg$ in the Galactic halo.  The radial velocity gradient along the stream is used to deduce transverse velocities as a function of distance, enabling a family of orbits to be computed.  We find that a direction of motion towards the Galactic disk coupled with a mid-stream distance of $\sim20\kpc$ provides a good match to the observed sky positions and radial velocities of the H$\I$ clouds comprising the stream.  With an estimated mass of $10^5\Msun$, its progenitor is likely to be a dwarf galaxy.  However, no stellar counterpart has been found amongst the currently known Galactic dwarf spheroidal galaxies or stellar streams and the exact origin of the stream is therefore currently unknown.

\end{abstract}


\section{Introduction}

The study of neutral hydrogen (H$\I$) structures in the Milky Way dates back to the early 1950s, with the first detections of 21cm emission by Ewen \& Purcell (1951)\nocite{1951Natur.168..356E} having followed the prediction by van der Hulst\nocite{Hulst1945} that this line should be detectable astronomically. Since these initial days of radio astronomy, the Galactic H$\I$ sky has been mapped repeatedly, providing very detailed maps of the H$\I$ distribution in the Milky Way as projected on the sky (Leiden-Dwingeloo Survey: \citealt{1997agnh.book.....H}, H$\I$ Parkes All Sky Survey: \citealt{2001MNRAS.322..486B}, Instituto Argentino de Radioastronom\'ia Survey: \citealt{2005A&A...440..767B}, Southern Galactic Plane Survey: \citealt{2005ApJS..158..178M}, Arecibo Galactic H$\I$ Survey: \citealt{2006ApJ...653.1210S}).  However, detections of gaseous structures in themselves do not, unfortunately, provide us with information about their three-dimensional spatial distribution.  In the disk plane, it is possible to de-project the sky-projected distribution from knowledge of their radial velocities and by assuming rotation around the Milky Way \citep[e.g. for the Galactic spiral arm structure;][]{2006Sci...312.1773L}.  In most cases, however, pinning down the location of a gaseous structure requires the knowledge of a stellar population that it can be associated with.

High-velocity clouds ($\HVC$s) of neutral hydrogen are a class of H$\I$ structures whose kinematic behaviour cannot be reconciled with that expected for disk-plane material.  While several theories exist for their origin \citep[the Galactic fountain, stripping from dwarf galaxies, remnants of galaxy formation; see e.g.][]{2004ASSL..312..341B}, none of these have outright popularity over others, primarily due to the fact that $\HVC$s must be treated on a case-by-case basis.  When deducible, their distances and metallicities can help to constrain their origin, but the former is particularly difficult to obtain for many of the Galactic $\HVC$s; their locations at often high Galactic latitudes means that identifying and utilising a halo field star to apply the interstellar absorption line technique, from which distance limits can be derived, is usually a taxing task.

In the context of stripped gas associated with dwarf galaxies, the Sagittarius dwarf galaxy is the only Milky Way satellite, aside from the Magellanic Clouds, for which there is an H$\I$ feature that could have been stripped from the main body of the dwarf galaxy \citep{2004ApJ...603L..77P}.   Indeed, the $\HVC$s comprising the Magellanic stream are the only ones for which there is definitive knowledge that the gas originated from the Milky Way satellite companions themselves.  In terms of satellite galaxies in the Local Group, the Magellanic system is unique in the sense that the two galaxies have clearly been interacting with one another as well as with the Milky Way; the binary interaction has most probably been crucial in producing the gaseous stream of over $100\deg$ in length. There are two reasons why we might expect to see more examples of H$\I$ streams than this one system that is currently known.  The first is that studies derived from cosmological simulations indicate that a significant fraction of dark matter subhalos fall into their more massive host halo in groups \citep{2008MNRAS.385.1365L}.  In this scenario, dwarf galaxies that are hosted within such subhalos will naturally have companions that may be involved in close interactions, depending on the degree of binding within the infalling group.  The second reason for expecting halo H$\I$ streams is that dwarf galaxies venturing too close to the parent galaxy should have their gas content at least partially stripped through tidal forces and/or ram pressure.  Indeed, the H$\I$ content of Milky Way and M31 satellites are well known to correlate with distance from their respective parent galaxies \citep{2009ApJ...696..385G}.\\

In this letter, we highlight the possibility that a collection of clouds belonging to an $\HVC$ complex known as GCN may be part of yet another gaseous stream at an average heliocentric distance of $\sim20\kpc$. In Section~\ref{sec:velocity}, we first outline how the velocity of the stream can be calculated as a function of distance. Section~\ref{sec:orbit} presents the orbit derived from this analysis and the results are summarised in Section~\ref{sec:summary}.

\section{Deriving the transverse velocity of a stream}
\label{sec:velocity}

Before turning to the analysis of the H$\I$ stream, we recall the technique for deducing the transverse velocity at a given location along a stream by using the gradient in line-of-sight velocity there, as has been described by \cite{2007MNRAS.378L..64J} and \cite{2008MNRAS.386L..47B}.  Its strength lies in the use of line-of-sight velocity data along an extended structure, so that proper motion measurements are not required to constrain an orbit, if at least one distance is known.  If no distance information is available, then the transverse velocities are calculated as distance-dependent quantities that can then be used to derive a family of possible orbits in a given gravitational potential.  Depending on the morphology of the orbit, both the distance of the stream and its direction of motion can be constrained by comparing the resulting orbit with observational data.

The formalism for deducing the component of the velocity vector transverse to the line of sight at a given point on a stream was provided in a previous paper \citep{2007MNRAS.378L..64J}.  Here, we simply quote the solution to the resulting quadratic equation for this transverse velocity, $v_s$:
\beqnarray                                                                      
\label{eqn:vs}                                                                  
v_s = \frac{1}{2} \left( \ud v_l/\ud\chi \pm \sqrt{\left(\ud v_l/\ud\chi\right)^2 - 4\left(\norml.\nabla\psi\right)d}\right)\,,
\eeqnarray         
where $v_l$ is the line-of-sight velocity corrected to the Galactic Standard of Rest (GSR), $\norml$ is the normalised line-of-sight vector, $d$ is the heliocentric distance, $\psi$ is the Galactic potential and $\chi$ is the angle along the stream, measured from some fiducial point.  The full velocity vector at any point along a stream is given by $\vecv = v_l\norml + v_s\norms$, where $\norms$ is the unit vector in the plane of the sky, along the apparent direction of motion of the stream at $\norml$; one possible method by which $\norms$ may be calculated is given by equation (16) in \cite{2008MNRAS.383.1686J}.  The two solutions for $v_s$ usually provide orbits in opposite senses.  Here, a comparison of the computed orbits with the H$\I$ data turns out to be sufficient to specify the direction of motion, given the large angular extent of the gaseous structure considered.

In the following analysis, we use a three-component Galactic model described by \cite{1990ApJ...348..485P}, which combines a Miyamoto-Nagai model for the disk and spheroid \citep{1975PASJ...27..533M}, and a near-logarithmic, spherical potential for the halo.  We also briefly describe the effect of changing the halo potential in Section~\ref{subsec:orbit}.

\section{Results}
\label{sec:orbit}

\subsection{Determining the orbit of GCN}
\label{subsec:orbit}

\begin{figure}
\begin{center}
\includegraphics[width=0.30\textwidth,angle=270]{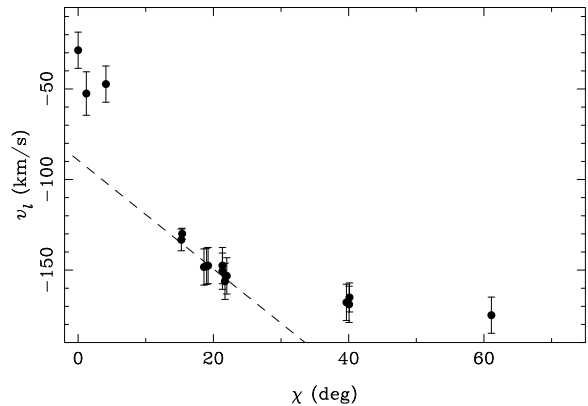}
\caption{Variation in line-of-sight velocity along the $\HVC$ complex GCN, where $\chi$ is the angle along the stream as measured from the point of lowest latitude and increases towards the Galactic plane.  The error bars on the velocities indicate the width of the velocity peak in the LAB data.  The dashed line indicates the velocity gradient at the starting point for the orbital integration.
\label{fig:chi_vl}}
\end{center}
\end{figure}

\begin{figure}
\begin{center}
\includegraphics[width=0.48\textwidth,angle=0]{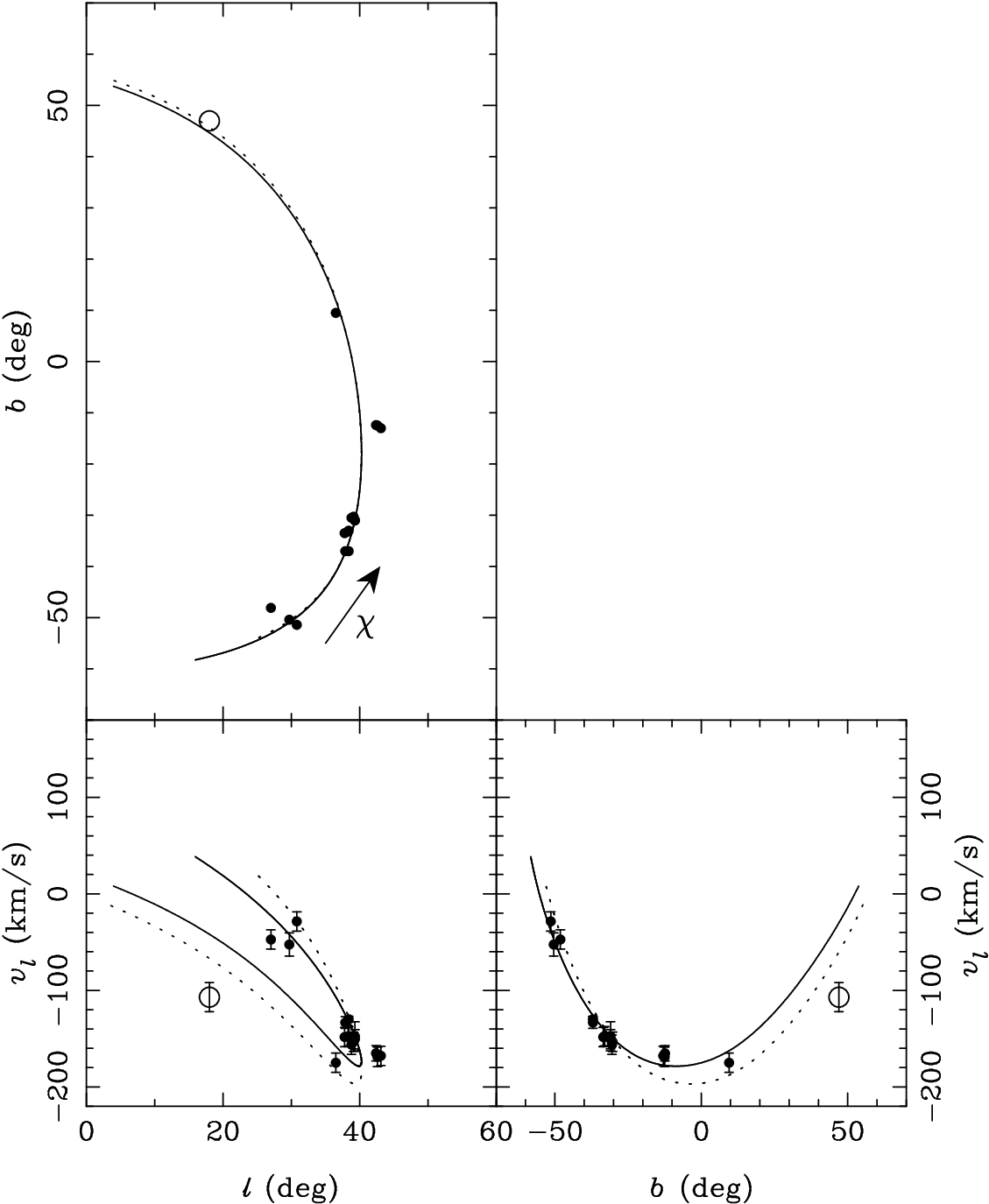} 
\caption{Orbital evolution in Galactic longitude and latitude $(\ell,b)$ and line-of-sight velocity relative to the Galactic Standard of Rest ($v_l$).  The panels provide a comparison of GCN data (dots) with an orbit (solid line) computed using parameters derived from its properties at $\ell = 39.3\deg$, $b=-31.0\deg$ (see main text for details).  The computed orbit places the H$\I$ complex at $25\kpc$ at the $\IC$ point. The open circle represents the $\CHVC$ at $b=47\deg$ (see discussion in Section \ref{subsec:assoc}).  Note that this solid-line orbit does {\it{not}} take this point into account; its inclusion here serves merely to illustrate the position of the $\CHVC$ relative to the stream's orbit. The dotted line describes an orbit that is constrained to have the same IC point as the orbit shown by the solid line, but passes closer to the $\CHVC$ in $\ell$, $b$ and $v_l$. \label{fig:lb_vl}}
\end{center}
\end{figure}

\begin{figure}
\begin{center}
\includegraphics[width=0.28\textwidth,angle=270]{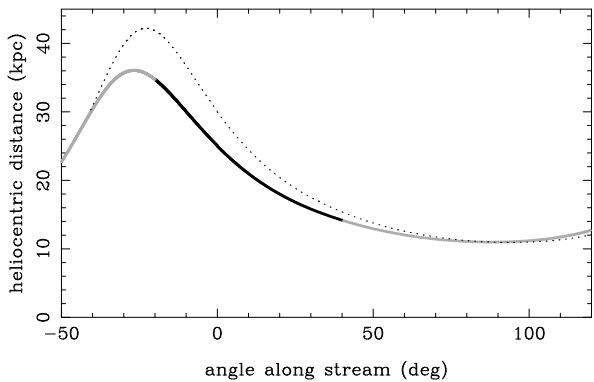}
\caption{Variation of heliocentric distance with angle along the stream (measured from the $\IC$ point), for the orbits shown in Figure~\ref{fig:lb_vl}. The solid line describes the case in which the $\CHVC$ is not included and corresponds to the orbit shown in Figure~\ref{fig:all-sky}, while the dotted line is with its inclusion.  The central region of the solid line, in black, corresponds to the extent of GCN that we have followed in the LAB data.  \label{fig:d_chi}}
\end{center}
\end{figure}

\begin{figure*}
\begin{center}
\includegraphics[width=0.50\textwidth,angle=270]{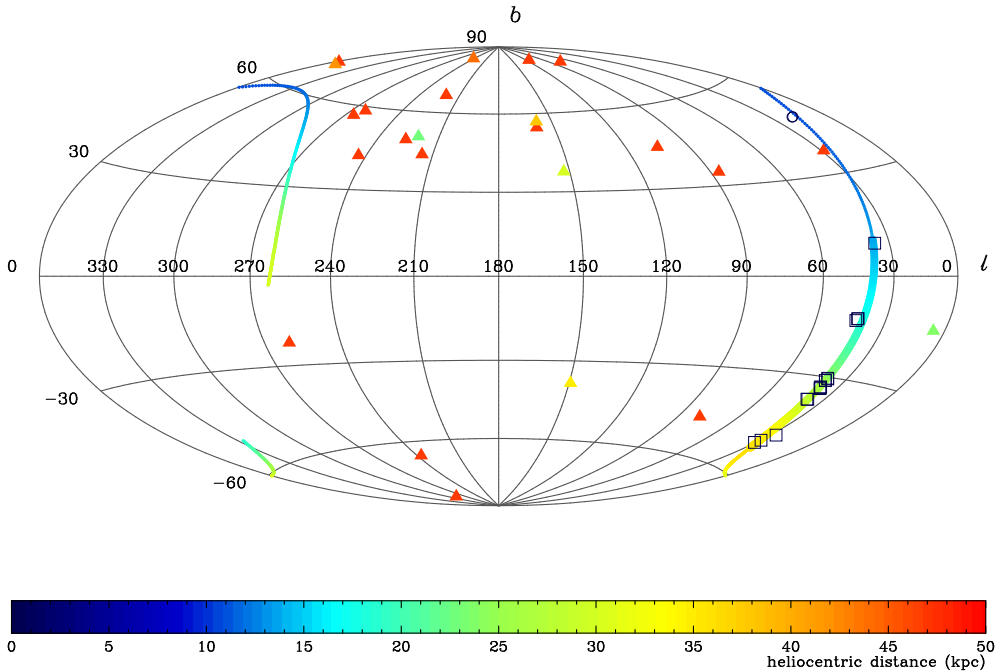}
\caption{Distance evolution of the derived orbit for GCN (corresponding to the orbit indicated by solid lines in Figures~\ref{fig:lb_vl} and \ref{fig:d_chi}), shown in Hammer projection.  The open squares denote positions of the H$\I$ clouds comprising GCN, while the open circle represents a compact high-velocity cloud that may or may not be associated with the main stream.  Triangles indicate the locations of Milky Way dwarf spheroidal galaxies, as listed in Table~\ref{tab:dwarfs}. \label{fig:all-sky}}
\end{center}
\end{figure*}

\begin{table}
\begin{center}
\begin{tabular}{lrrrrr}
\hline
\multicolumn{1}{c}{}% satellite name
& \multicolumn{1}{c}{$\ell$}
& \multicolumn{1}{c}{$b$}
& \multicolumn{1}{c}{$v_\odot$}
& \multicolumn{1}{c}{$v_l$}
& \multicolumn{1}{c}{$d$}\\
% units:
\multicolumn{1}{c}{}
& \multicolumn{1}{c}{$(\deg)$}
& \multicolumn{1}{c}{$(\deg)$}
& \multicolumn{1}{c}{(km/s)}
& \multicolumn{1}{c}{(km/s)}
& \multicolumn{1}{c}{(kpc)}\\
\hline
 Bo\"otes I        &  358.1  &  69.6   &   99   &  106   &   66  \\
 Bo\"otes II       &  353.7  &  68.9   & $-$117 & $-$116 &   42  \\
 Canes Venatici I  &   74.3  &  79.8   &   31   &   78   &  218  \\
 Canes Venatici II &  113.6  &  82.7   & $-$129 & $-$96  &  160  \\
 Carina            &  260.1  & $-$22.2 &  224   &    8   &  101  \\
 Coma Berenices    &  241.9  &  83.6   &   98   &   82   &   44  \\
 Draco             &   86.4  &  34.7   & $-$293 & $-$98  &   82  \\
 Fornax            &  237.1  & $-$65.7 &   53   & $-$36  &  138  \\
 Hercules          &   28.7  &  36.9   &   45   &  145   &  132  \\
 Leo I             &  226.0  &  49.1   &  286   &  178   &  250  \\
 Leo II            &  220.2  &  67.2   &   76   &   22   &  205  \\
 Leo IV            &  265.4  &  56.5   &  132   &   10   &  160  \\
 Leo V             &  261.9  &  58.5   &  173   &   58   &  178  \\
 Leo T             &  214.9  &  43.7   &   38   & $-$58  &  407  \\
Pisces II         & 79.2 & $-$47.1 & $\cdots$ & $\cdots$ & 180 \\
 Sagittarius       &    5.6  & $-$14.1 &  140   &  169   &   24  \\
 Sculptor          &  287.5  & $-$83.2 &  108   &   75   &   79  \\
 Segue 1           &  220.5  &  50.4   &  206   &  111   &   23  \\
 Segue 2           &  149.4  & $-$38.1 & $-$39  &   44   &   35  \\
 Sextans           &  243.5  &  42.3   &  227   &   75   &   86  \\
 Ursa Major I      &  159.4  &  54.4   & $-$55  & $-$7   &   97  \\
 Ursa Major II     &  152.5  &  37.4   & $-$116 & $-$33  &   30  \\
 Ursa Minor        &  105.0  &  44.8   & $-$248 & $-$86  &   66  \\
 Willman 1         &  158.6  &  56.8   & $-$12  &   36   &   38  \\
\hline
\end{tabular}
\caption{Properties of Milky Way dwarf spheroidal galaxies shown in Figure~\ref{fig:all-sky}.  The columns give the Galactic longitude and latitude, heliocentric radial velocity, line-of-sight velocity relative to the Galactic Standard of Rest and heliocentric distance.  Radial velocities and distances are taken from those compiled by \citet{Mateo1998} and \citet{2008ApJ...684.1075M}, with radial velocities for the `ultra-faint' dwarf galaxies taken from \citet{2007MNRAS.380..281M} and \citet{2007ApJ...670..313S}, with additional information for Segue~1, Leo~V, Segue~2 and Pisces~II from \citet{2009ApJ...692.1464G}, \citet{2008ApJ...686L..83B,2009MNRAS.397.1748B} and \citet{2010ApJ...712L.103B}, respectively.}
\label{tab:dwarfs}
\end{center}
\end{table}

Using the all-sky $\HVC$ map by \citet[][figure~1b]{2004ASSL..312...25W} as a visual guide, we extract several GCN H$\I$ features in the Leiden/Argentine/Bonn (LAB) Galactic H$\I$ survey data cube \citep{2005A&A...440..775K} lying along the general direction of $\ell\simeq33-40\deg$ with negative velocities relative to the GSR.\footnote{Another $\HVC$ complex (GCP) overlaps GCN spatially for a good chunk of the latter.  However, these two structures have rather contrasting line-of-sight velocities so that any physical association between them is clearly ruled out and their emission is easily separated in the LAB data.}  Figure~\ref{fig:chi_vl} shows the evolution of $v_l$ along the string of $\HVC$s that indicate a strong communal trend in both sky position and velocity, the data for which can then be used to calculate $\ud v_l/\ud\chi$ and hence the transverse velocity along the stream.

In defining the initial condition ($\IC$) point for the orbit calculation, it is important to choose a location on the stream where the behaviour of $\ud v_l/\ud\chi$ is well defined. This usually corresponds to a point near the middle of the stream's extent.  Here, the $\IC$ point chosen to satisfy these criteria is at a Galactic longitude and latitude of $(\ell,b) = (39.3,-31.0)\deg$, with the corresponding properties $v_l = -147.5\kms$ and $\ud v_l/\ud\chi = -3.0\kms/{\mathrm{deg}}$ (indicated by the dashed line in Figure~\ref{fig:chi_vl}).  The free parameter in this analysis is the heliocentric distance to the stream.  However, there is also some uncertainty in the velocity gradient along the stream of order $0.5\kms/{\mathrm{deg}}$; we therefore treat this gradient as a somewhat variable parameter.  In order to recover a good match to the position and velocity of the H$\I$ data points, we find that the distance to the $\IC$ point is reasonably well constrained at $\sim25\kpc$ with the remaining orbital parameters as stated above.\footnote{Altering this initial distance by $\pm10\kpc$ leads to orbits that are clearly discrepant with one half of the stream.}  The orbit that results from taking this set of parameters is one of the best found and is shown by the solid lines in Figure~\ref{fig:lb_vl}, where the three panels indicate the variation of each of the variables $\ell$, $b$ and $v_l$ as a function of the other variables.  Note that the stream mainly extends in Galactic latitude (in other words, $\chi$ is effectively given by $b$), and so the panel showing the trend of $v_l$ in longitude is less instructive in characterising and understanding the orbit than the other two panels.  The solid line in Figure~\ref{fig:d_chi} shows the evolution of heliocentric distance along this orbit, with the highlighted central section corresponding to the extent traceable in the LAB data.

We note that the details of any orbit calculated are naturally dependent on the choice of the Galactic potential.  As many, more cosmologically-motivated halo profiles than that described by \cite{1990ApJ...348..485P} now exist in the literature, the analysis was also re-performed with Paczy\'nski's halo being replaced by an adiabatically-contracted NFW \citep{NFW96,1997ApJ...490..493N} profile with parameters constrained by \citet{2008ApJ...684.1143X}, in order to check the dependency of the results on the choice of the halo potential.
Keeping the same initial conditions as before, we find that the change in halo potential leads only to small changes in the resulting orbit that are mainly manifest above the Galactic plane.  Within the region of sky covered by GCN, the differences in the results are negligible in both the radial velocity trend and sky projection of the orbit.  The distances are slightly affected by the fact that the Xue et al. halo contains less mass than Paczy\'nski's.    As a result, the orbit has a larger apocentre and reaches a smaller pericentre, with a maximum change in distance of $\sim10\%$ compared to the original orbit.  The qualitative results and most of the quantitative results are, however, unchanged by the modification of the halo potential.

\subsection{Checking for associations}
\label{subsec:assoc}

A different representation of the stream's orbit is provided in Figure~\ref{fig:all-sky}, where the `anti-centre projection' allows us to directly compare our results with the Wakker (2004) $\HVC$ map. The same orbit as that shown by the solid lines in Figures~\ref{fig:lb_vl} and \ref{fig:d_chi} is now overlaid with the H$\I$ stream data points as open squares; additionally, all of the known Milky Way dwarf spheroidal (dSph) galaxies are plotted as filled triangles (Table~\ref{tab:dwarfs}).  Although there are a few spatial coincidences, the distance and velocity of the orbit at these locations firmly rule out possible associations of this H$\I$ stream with any of the currently known Milky Way dSphs.  

The Milky Way is also host to $\sim150$ globular clusters \citep{1996AJ....112.1487H}.  Of these, a handful of globular clusters with latitudes between $-50\deg$ and $0\deg$ lie close to the stream but, with velocity differences of more than $100\kms$ from the GCN clouds or the inferred orbit, connections to the H$\I$ stream are unlikely.  The same conclusion holds for three other globular clusters at $40\deg < b < 50\deg$ that lie within $10\deg$ of the orbit there, although one should remember that the orbital path at these latitudes is already a strong extrapolation of the orbit derived for the main part of the H$\I$ stream residing at negative latitudes.

Thanks to the advent of large sky surveys, numerous detections of Galactic stellar streams have been reported in the literature in the last few years.  We therefore also considered the possibility that the GCN stream might have a counterpart in one of these recent discoveries, if not in the globular clusters or intact dwarf spheroidals.  We find that no known stellar stream [Monoceros stream \citep{2002ApJ...569..245N,2003MNRAS.340L..21I}, Orphan stream \citep{2006ApJ...645L..37G,2007ApJ...658..337B}; Grillmair-Dionatos stream \citep{2006ApJ...643L..17G,2009ApJ...697..207W}; Acheron, Cocytos, Lethe, and Styx streams \citep{2009ApJ...693.1118G}; solar neighbourhood streams in SDSS DR7 \citep{2009ApJ...698..865K}; Cetus polar stream \citep{2009ApJ...700L..61N}; Pisces overdensity \citep{2009MNRAS.398.1757W}] coincides with the orbit determined in this letter for the GCN stream.  Although inhabiting a similar region of the sky, the stream is also offset from the Magellanic system in location and velocity.  Combining this with our check for coincidences with the stellar streams therefore leads us to conclude that the H$\I$ stream has no currently known stellar counterpart.

Finally, we find that this orbit passes very close to a compact high-velocity cloud ($\CHVC$) in the Wakker $\HVC$ map at $(\ell,b) = (18,47)\deg$.  Perusal of the LAB data shows this $\CHVC$ to have a line-of-sight velocity of $(-107\pm15)\kms$.  As shown by the dotted lines in Figures~\ref{fig:lb_vl} and \ref{fig:d_chi}, its inclusion in our analysis helps to constrain the distance of the orbit further by requiring the orbit to trace as much as possible the clouds over a larger extent on the sky, but the association of this H$\I$ clump with the rest of the structure is unclear from the current data.  Higher resolution H$\I$ data\footnote{For example, very recent results from the Effelsberg-Bonn H$\I$ survey \citep{2010arXiv1007.3363W} clearly show head-tail features and filamentary structures in new, high resolution data for GCN clouds.} would be required to see whether this $\CHVC$ exhibits an elongation along the direction of motion expected from the orbit; naturally, detection of lower column density material along this direction would also aid in solidifying this hypothesis.

\subsection{Discussion}
\label{subsec:discussion}

Despite its `orphan' nature in having no obvious progenitor or stellar counterpart, our findings provide evidence for the first halo H$\I$ stream outside of the Magellanic stream at distances where it is conceivable for H$\I$ structures to display relatively ballistic motion.  Although the orbits were always integrated for several wraps, here we have shown only the first $660\Myr$ to either side of the $\IC$ point.  The reason for this is three fold: the first is simply a matter of clarity, so that the reader does not have to untangle an array of different orbital wraps on such a map.  The second is that the exact form of the gravitational potential is, as always, somewhat uncertain.  In addition, we do not expect the motion of any halo stream (whether stellar or gaseous) to follow an exact orbit, as the extended profile itself is created through changes in orbital energy and angular momentum amongst the stream's constituents.  This is even more important for gaseous structures than it is for stellar streams, as gas is subject to non-gravitational influences in the form of ram pressure and drag forces, effects that do not concern a stellar stream.  We therefore do not expect to find a perfect orbital match and, rather, content ourselves with an orbit that reproduces well the general trend in position and velocity of the string of H$\I$ clouds of our interest.

As we were only able to identify a few of the brightest regions in the LAB data, it is difficult to make an accurate mass estimate of the stream.  We therefore simply use a representative column density of $10^{19}\cm2$ \citep{2004ASSL..312...25W} and a total angular size of 20~${\mathrm{deg}}^2$ at an average distance of $20\kpc$ to arrive at a conservative lower mass estimate of $2\times10^{5}\Msun$.  Studies of the H$\I$ content of Local Group dwarf galaxies \citep{2009ApJ...696..385G} indicate this to be a plausible gas content for an undisturbed, distant dwarf galaxy.  Observing such a quantity of H$\I$ spread into a stream would, however, clearly be more challenging than observing it as a bound and more concentrated structure, and may explain why the stream had not been identified before.  It is also not surprising that the GCN stream would be much less massive than the Magellanic stream given the masses of its progenitors, the Large and Small Magellanic Clouds, which themselves are a few orders of magnitude more massive than most of the dwarf companions of the Milky Way.

\section{Summary}
\label{sec:summary}

We have shown that a string of H$\I$ clouds that form part of the high-velocity cloud complex known as GCN is a probable gaseous stream in the Galactic halo at a distance of $\sim20\kpc$.  We determine its orbit by utilising the large gradient in line-of-sight velocity along its extent, and determine its direction of motion to be towards the Galactic plane.  We also identify a compact $\HVC$, whose sky position nearly intersects with the forward projection of the stream's orbit, and whose velocity differs by approximately $50\kms$ from that of the orbit there.  This association, however, remains a speculative one before further H$\I$ studies along the direction of the orbit presented here can be performed.  The origin of such a gaseous stream with an estimated gas content of a few times $10^5\Msun$ is most probably a satellite galaxy of the Milky Way.  However, no progenitor has been found amongst the currently known Galactic dSphs or globular clusters through a comparison of their locations and velocities with the stream's orbit, hence the exact origin of the stream is currently unknown.

\subsection*{Acknowledgements} 

I would like to thank Donald Lynden-Bell for many inspirational and insightful discussions that encouraged me to pursue this work. I am also grateful to Nicolas Martin for useful discussions and a careful reading of the manuscript, and to the anonymous referee for helpful comments and suggestions.

\bibliographystyle{mn2e}

\begin{thebibliography}{}

\bibitem[\protect\citeauthoryear{{Bajaja}, {Arnal}, {Larrarte}, {Morras},
  {P{\"o}ppel} \& {Kalberla}}{{Bajaja} et~al.}{2005}]{2005A&A...440..767B}
{Bajaja} E.,  {Arnal} E.~M.,  {Larrarte} J.~J.,  {Morras} R.,  {P{\"o}ppel}
  W.~G.~L.,    {Kalberla} P.~M.~W.,  2005, \aap, 440, 767

\bibitem[\protect\citeauthoryear{{Barnes}, {Staveley-Smith}, {de Blok},
  {Oosterloo}, {Stewart}, {Wright}, {Banks}, {Bhathal}, {Boyce}, {Calabretta},
  {Disney}, {Drinkwater}, {Ekers}, {Freeman}, {Gibson}, {Green}, {Haynes} \&
  {et al.}}{{Barnes} et~al.}{2001}]{2001MNRAS.322..486B}
{Barnes} D.~G.,  {Staveley-Smith} L.,  {de Blok} W.~J.~G.,  {Oosterloo} T.,
  {Stewart} I.~M.,  {Wright} A.~E.,  {Banks} G.~D.,  {Bhathal} R.,  {Boyce}
  P.~J.,  {Calabretta} M.~R.,  {Disney} M.~J.,  {Drinkwater} M.~J.,  {Ekers}
  R.~D.,  {Freeman} K.~C.,  {Gibson} B.~K.,  {Green} A.~J.,  {Haynes} R.~F.,
  {et al.} 2001, \mnras, 322, 486

\bibitem[\protect\citeauthoryear{{Belokurov}, {Evans}, {Irwin}, {Lynden-Bell},
  {Yanny}, {Vidrih}, {Gilmore}, {Seabroke} \& {et al.}}{{Belokurov}
  et~al.}{2007}]{2007ApJ...658..337B}
{Belokurov} V.,  {Evans} N.~W.,  {Irwin} M.~J.,  {Lynden-Bell} D.,  {Yanny} B.,
   {Vidrih} S.,  {Gilmore} G.,  {Seabroke} G.,    {et al.} 2007, \apj, 658, 337

\bibitem[\protect\citeauthoryear{{Belokurov}, {Walker}, {Evans}, {Faria},
  {Gilmore}, {Irwin}, {Koposov}, {Mateo}, {Olszewski} \& {Zucker}}{{Belokurov}
  et~al.}{2008}]{2008ApJ...686L..83B}
{Belokurov} V.,  {Walker} M.~G.,  {Evans} N.~W.,  {Faria} D.~C.,  {Gilmore} G.,
   {Irwin} M.~J.,  {Koposov} S.,  {Mateo} M.,  {Olszewski} E.,    {Zucker}
  D.~B.,  2008, \apjl, 686, L83

\bibitem[\protect\citeauthoryear{{Belokurov}, {Walker}, {Evans}, {Gilmore},
  {Irwin}, {Just}, {Koposov}, {Mateo}, {Olszewski}, {Watkins} \&
  {Wyrzykowski}}{{Belokurov} et~al.}{2010}]{2010ApJ...712L.103B}
{Belokurov} V.,  {Walker} M.~G.,  {Evans} N.~W.,  {Gilmore} G.,  {Irwin} M.~J.,
   {Just} D.,  {Koposov} S.,  {Mateo} M.,  {Olszewski} E.,  {Watkins} L.,
  {Wyrzykowski} L.,  2010, \apjl, 712, L103

\bibitem[\protect\citeauthoryear{{Belokurov}, {Walker}, {Evans}, {Gilmore},
  {Irwin}, {Mateo}, {Mayer}, {Olszewski}, {Bechtold} \&
  {Pickering}}{{Belokurov} et~al.}{2009}]{2009MNRAS.397.1748B}
{Belokurov} V.,  {Walker} M.~G.,  {Evans} N.~W.,  {Gilmore} G.,  {Irwin} M.~J.,
   {Mateo} M.,  {Mayer} L.,  {Olszewski} E.,  {Bechtold} J.,    {Pickering} T.,
   2009, \mnras, 397, 1748

\bibitem[\protect\citeauthoryear{{Binney}}{{Binney}}{2008}]{2008MNRAS.386L..47%
B}
{Binney} J.,  2008, \mnras, 386, L47

\bibitem[\protect\citeauthoryear{{Bregman}}{{Bregman}}{2004}]{2004ASSL..312..3%
41B}
{Bregman} J.~N.,  2004, in {H.~van Woerden, B.~P.~Wakker, U.~J.~Schwarz, \&
  K.~S.~de Boer } ed., High Velocity Clouds Vol.~312 of Astrophysics and Space
  Science Library, {The Origin of the High-Velocity Clouds}.
p.~341

\bibitem[\protect\citeauthoryear{{Ewen} \& {Purcell}}{{Ewen} \&
  {Purcell}}{1951}]{1951Natur.168..356E}
{Ewen} H.~I.,  {Purcell} E.~M.,  1951, \nat, 168, 356

\bibitem[\protect\citeauthoryear{{Geha}, {Willman}, {Simon}, {Strigari},
  {Kirby}, {Law} \& {Strader}}{{Geha} et~al.}{2009}]{2009ApJ...692.1464G}
{Geha} M.,  {Willman} B.,  {Simon} J.~D.,  {Strigari} L.~E.,  {Kirby} E.~N.,
  {Law} D.~R.,    {Strader} J.,  2009, \apj, 692, 1464

\bibitem[\protect\citeauthoryear{{Grcevich} \& {Putman}}{{Grcevich} \&
  {Putman}}{2009}]{2009ApJ...696..385G}
{Grcevich} J.,  {Putman} M.~E.,  2009, \apj, 696, 385

\bibitem[\protect\citeauthoryear{{Grillmair}}{{Grillmair}}{2006}]{2006ApJ...64%
5L..37G}
{Grillmair} C.~J.,  2006, \apjl, 645, L37

\bibitem[\protect\citeauthoryear{{Grillmair}}{{Grillmair}}{2009}]{2009ApJ...69%
3.1118G}
{Grillmair} C.~J.,  2009, \apj, 693, 1118

\bibitem[\protect\citeauthoryear{{Grillmair} \& {Dionatos}}{{Grillmair} \&
  {Dionatos}}{2006}]{2006ApJ...643L..17G}
{Grillmair} C.~J.,  {Dionatos} O.,  2006, \apjl, 643, L17

\bibitem[\protect\citeauthoryear{{Harris}}{{Harris}}{1996}]{1996AJ....112.1487%
H}
{Harris} W.~E.,  1996, \aj, 112, 1487

\bibitem[\protect\citeauthoryear{{Hartmann} \& {Burton}}{{Hartmann} \&
  {Burton}}{1997}]{1997agnh.book.....H}
{Hartmann} D.,  {Burton} W.~B.,  1997, {Atlas of Galactic neutral hydrogen}.
Cambridge; New York: Cambridge University Press, ISBN 0521471117

\bibitem[\protect\citeauthoryear{{Ibata}, {Irwin}, {Lewis}, {Ferguson} \&
  {Tanvir}}{{Ibata} et~al.}{2003}]{2003MNRAS.340L..21I}
{Ibata} R.~A.,  {Irwin} M.~J.,  {Lewis} G.~F.,  {Ferguson} A.~M.~N.,
  {Tanvir} N.,  2003, \mnras, 340, L21

\bibitem[\protect\citeauthoryear{{Jin} \& {Lynden-Bell}}{{Jin} \&
  {Lynden-Bell}}{2007}]{2007MNRAS.378L..64J}
{Jin} S.,  {Lynden-Bell} D.,  2007, \mnras, 378, L64

\bibitem[\protect\citeauthoryear{{Jin} \& {Lynden-Bell}}{{Jin} \&
  {Lynden-Bell}}{2008}]{2008MNRAS.383.1686J}
{Jin} S.,  {Lynden-Bell} D.,  2008, \mnras, 383, 1686

\bibitem[\protect\citeauthoryear{{Kalberla}, {Burton}, {Hartmann}, {Arnal},
  {Bajaja}, {Morras} \& {P{\"o}ppel}}{{Kalberla}
  et~al.}{2005}]{2005A&A...440..775K}
{Kalberla} P.~M.~W.,  {Burton} W.~B.,  {Hartmann} D.,  {Arnal} E.~M.,  {Bajaja}
  E.,  {Morras} R.,    {P{\"o}ppel} W.~G.~L.,  2005, \aap, 440, 775

\bibitem[\protect\citeauthoryear{{Klement}, {Rix}, {Flynn}, {Fuchs}, {Beers},
  {Allende Prieto}, {Bizyaev}, {Brewington}, S., {Malanushenko},
  {Malanushenko}, {Oravetz}, {Pan}, {Re Fiorentin}, {Simmons} \&
  {Snedden}}{{Klement} et~al.}{2009}]{2009ApJ...698..865K}
{Klement} R.,  {Rix} H.-W.,  {Flynn} C.,  {Fuchs} B.,  {Beers} T.~C.,  {Allende
  Prieto} C.,  {Bizyaev} D.,  {Brewington} H.,  S. L.~Y.,  {Malanushenko} E.,
  {Malanushenko} V.,  {Oravetz} D.,  {Pan} K.,  {Re Fiorentin} P.,  {Simmons}
  A.,    {Snedden} S.,  2009, \apj, 698, 865

\bibitem[\protect\citeauthoryear{{Levine}, {Blitz} \& {Heiles}}{{Levine}
  et~al.}{2006}]{2006Sci...312.1773L}
{Levine} E.~S.,  {Blitz} L.,    {Heiles} C.,  2006, Science, 312, 1773

\bibitem[\protect\citeauthoryear{{Li} \& {Helmi}}{{Li} \&
  {Helmi}}{2008}]{2008MNRAS.385.1365L}
{Li} Y.-S.,  {Helmi} A.,  2008, \mnras, 385, 1365

\bibitem[\protect\citeauthoryear{{Martin}, {de Jong} \& {Rix}}{{Martin}
  et~al.}{2008}]{2008ApJ...684.1075M}
{Martin} N.~F.,  {de Jong} J.~T.~A.,    {Rix} H.-W.,  2008, \apj, 684, 1075

\bibitem[\protect\citeauthoryear{{Martin}, {Ibata}, {Chapman}, {Irwin} \&
  {Lewis}}{{Martin} et~al.}{2007}]{2007MNRAS.380..281M}
{Martin} N.~F.,  {Ibata} R.~A.,  {Chapman} S.~C.,  {Irwin} M.,    {Lewis}
  G.~F.,  2007, \mnras, 380, 281

\bibitem[\protect\citeauthoryear{{Mateo}}{{Mateo}}{1998}]{Mateo1998}
{Mateo} M.~L.,  1998, \araa, 36, 435

\bibitem[\protect\citeauthoryear{{McClure-Griffiths}, {Dickey}, {Gaensler},
  {Green}, {Haverkorn} \& {Strasser}}{{McClure-Griffiths}
  et~al.}{2005}]{2005ApJS..158..178M}
{McClure-Griffiths} N.~M.,  {Dickey} J.~M.,  {Gaensler} B.~M.,  {Green} A.~J.,
  {Haverkorn} M.,    {Strasser} S.,  2005, \apjs, 158, 178

\bibitem[\protect\citeauthoryear{{Miyamoto} \& {Nagai}}{{Miyamoto} \&
  {Nagai}}{1975}]{1975PASJ...27..533M}
{Miyamoto} M.,  {Nagai} R.,  1975, \pasj, 27, 533

\bibitem[\protect\citeauthoryear{{Navarro}, {Frenk} \& {White}}{{Navarro}
  et~al.}{1996}]{NFW96}
{Navarro} J.~F.,  {Frenk} C.~S.,    {White} S.~D.~M.,  1996, \apj, 462, 563

\bibitem[\protect\citeauthoryear{{Navarro}, {Frenk} \& {White}}{{Navarro}
  et~al.}{1997}]{1997ApJ...490..493N}
{Navarro} J.~F.,  {Frenk} C.~S.,    {White} S.~D.~M.,  1997, \apj, 490, 493

\bibitem[\protect\citeauthoryear{{Newberg}, {Yanny}, {Rockosi}, {Grebel},
  {Rix}, {Brinkmann}, {Csabai}, {Hennessy}, {Hindsley}, {Ibata}, {Ivezi{\'c}},
  {Lamb}, {Nash}, {Odenkirchen}, {Rave}, {Schneider}, {Smith}, {Stolte} \&
  {York}}{{Newberg} et~al.}{2002}]{2002ApJ...569..245N}
{Newberg} H.~J.,  {Yanny} B.,  {Rockosi} C.,  {Grebel} E.~K.,  {Rix} H.-W.,
  {Brinkmann} J.,  {Csabai} I.,  {Hennessy} G.,  {Hindsley} R.~B.,  {Ibata} R.,
   {Ivezi{\'c}} Z.,  {Lamb} D.,  {Nash} E.~T.,  {Odenkirchen} M.,  {Rave}
  H.~A.,  {Schneider} D.~P.,  {Smith} J.~A.,  {Stolte} A.,    {York} D.~G.,
  2002, \apj, 569, 245

\bibitem[\protect\citeauthoryear{{Newberg}, {Yanny} \& {Willett}}{{Newberg}
  et~al.}{2009}]{2009ApJ...700L..61N}
{Newberg} H.~J.,  {Yanny} B.,    {Willett} B.~A.,  2009, \apjl, 700, L61

\bibitem[\protect\citeauthoryear{{Paczy\'nski}}{{Paczy\'nski}}{1990}]{1990ApJ.%
..348..485P}
{Paczy\'nski} B.,  1990, \apj, 348, 485

\bibitem[\protect\citeauthoryear{{Putman}, {Thom}, {Gibson} \&
  {Staveley-Smith}}{{Putman} et~al.}{2004}]{2004ApJ...603L..77P}
{Putman} M.~E.,  {Thom} C.,  {Gibson} B.~K.,    {Staveley-Smith} L.,  2004,
  \apjl, 603, L77

\bibitem[\protect\citeauthoryear{{Simon} \& {Geha}}{{Simon} \&
  {Geha}}{2007}]{2007ApJ...670..313S}
{Simon} J.~D.,  {Geha} M.,  2007, \apj, 670, 313

\bibitem[\protect\citeauthoryear{{Stanimirovi{\'c}}, {Putman}, {Heiles},
  {Peek}, {Goldsmith}, {Koo}, {Kr{\v c}o}, {Lee}, {Mock}, {Muller}, {Pandian},
  {Parsons}, {Tang} \& {Werthimer}}{{Stanimirovi{\'c}}
  et~al.}{2006}]{2006ApJ...653.1210S}
{Stanimirovi{\'c}} S.,  {Putman} M.,  {Heiles} C.,  {Peek} J.~E.~G.,
  {Goldsmith} P.~F.,  {Koo} B.-C.,  {Kr{\v c}o} M.,  {Lee} J.-J.,  {Mock} J.,
  {Muller} E.,  {Pandian} J.~D.,  {Parsons} A.,  {Tang} Y.,    {Werthimer} D.,
  2006, \apj, 653, 1210

\bibitem[\protect\citeauthoryear{{van de Hulst}}{{van de
  Hulst}}{1945}]{Hulst1945}
{van de Hulst} H.~C.,  1945, Ned. tijd. natuurkunde, 11, 210

\bibitem[\protect\citeauthoryear{{Wakker}}{{Wakker}}{2004}]{2004ASSL..312...25%
W}
{Wakker} B.~P.,  2004, in {van Woerden} H.,  {Wakker} B.~P.,  {Schwarz} U.~J.,
   {de Boer} K.~S.,  eds, High Velocity Clouds Vol.~312 of Astrophysics and
  Space Science Library, {HVC/IVC Maps and HVC Distribution Functions}.
p.~25

\bibitem[\protect\citeauthoryear{{Watkins}, {Evans}, {Belokurov}, {Smith},
  {Hewett}, {Bramich}, {Gilmore}, {Irwin}, {Vidrih}, {Wyrzykowski} \&
  {Zucker}}{{Watkins} et~al.}{2009}]{2009MNRAS.398.1757W}
{Watkins} L.~L.,  {Evans} N.~W.,  {Belokurov} V.,  {Smith} M.~C.,  {Hewett}
  P.~C.,  {Bramich} D.~M.,  {Gilmore} G.~F.,  {Irwin} M.~J.,  {Vidrih} S.,
  {Wyrzykowski} {\L}.,    {Zucker} D.~B.,  2009, \mnras, 398, 1757

\bibitem[\protect\citeauthoryear{{Willett}, {Newberg}, {Zhang}, {Yanny} \&
  {Beers}}{{Willett} et~al.}{2009}]{2009ApJ...697..207W}
{Willett} B.~A.,  {Newberg} H.~J.,  {Zhang} H.,  {Yanny} B.,    {Beers} T.~C.,
  2009, \apj, 697, 207

\bibitem[\protect\citeauthoryear{{Winkel}, {Kerp}, {Kalberla} \& {Ben
  Bekhti}}{{Winkel} et~al.}{2010}]{2010arXiv1007.3363W}
{Winkel} B.,  {Kerp} J.,  {Kalberla} P.~M.~W.,    {Ben Bekhti} N.,  2010,
  arXiv:1007.3363

\bibitem[\protect\citeauthoryear{{Xue}, {Rix}, {Zhao}, {Re Fiorentin}, {Naab},
  {Steinmetz}, {van den Bosch}, {Beers}, {Lee}, {Bell}, {Rockosi}, {Yanny},
  {Newberg}, {Wilhelm}, {Kang}, {Smith} \& {Schneider}}{{Xue}
  et~al.}{2008}]{2008ApJ...684.1143X}
{Xue} X.~X.,  {Rix} H.~W.,  {Zhao} G.,  {Re Fiorentin} P.,  {Naab} T.,
  {Steinmetz} M.,  {van den Bosch} F.~C.,  {Beers} T.~C.,  {Lee} Y.~S.,  {Bell}
  E.~F.,  {Rockosi} C.,  {Yanny} B.,  {Newberg} H.,  {Wilhelm} R.,  {Kang} X.,
  {Smith} M.~C.,    {Schneider} D.~P.,  2008, \apj, 684, 1143

\end{thebibliography}

\end{document}